\documentclass[conference]{IEEEtran}
\IEEEoverridecommandlockouts

\usepackage{amsmath}
\usepackage{amssymb}
\usepackage{mathtools}
\usepackage{amsthm}

\usepackage{url}

\usepackage{setspace}
\usepackage{bm}
\usepackage{tikz}
\usepackage{braket}
\usepackage{amsthm}
\usepackage{lipsum}
\usepackage{subfig}
\usepackage{xcolor}
\usepackage{wrapfig}
\usepackage{amsmath}
\usepackage{balance}
\usepackage{environ}
\usepackage{physics}
\usepackage{colortbl}
\usepackage{enumitem}
\usepackage{booktabs}
\usepackage{graphicx}
\usepackage{textcomp}
\usepackage{algorithm}
\usepackage{algorithmic}
\usepackage[normalem]{ulem}
\usepackage{soul}
\usepackage{tcolorbox,tikz}
\usepackage[utf8]{inputenc} 
\usepackage[T1]{fontenc}    
\usepackage{hyperref}       
\usepackage{url}            
\usepackage{booktabs}       
\usepackage{amsfonts}       
\usepackage{nicefrac}       
\usepackage{microtype}      
\usepackage{xcolor}         

\newcommand{\rev}[1]{\textcolor{black}{#1}}

\theoremstyle{plain}
\newtheorem{theorem}{Theorem}[section]

\newtheorem{lemma}[theorem]{Lemma}

\theoremstyle{definition}
\newtheorem{definition}[theorem]{Definition}
\newtheorem{assumption}[theorem]{Assumption}
\theoremstyle{remark}
\newtheorem{remark}[theorem]{Remark}

\newcommand{\sol}{CircuitTree}

\def\BibTeX{{\rm B\kern-.05em{\sc i\kern-.025em b}\kern-.08em
    T\kern-.1667em\lower.7ex\hbox{E}\kern-.125emX}}

\begin{document}

\date{}

\title{Domain-Aware Probability Sampling for Hybrid Quantum Systems using Bayesian Optimization}

\author{
\IEEEauthorblockN{Nicholas S. DiBrita, Jason Han}
\IEEEauthorblockA{\textit{Rice University}\\Houston, TX, USA}
\and
\IEEEauthorblockN{Krishna Bhatia}
\IEEEauthorblockA{\textit{Fractal Analytics}\\Mumbai, MH, India}
\and
\IEEEauthorblockN{Younghyun Cho}
\IEEEauthorblockA{\textit{Santa Clara University}\\Santa Clara, CA, USA}
\and
\IEEEauthorblockN{Hengrui Luo, Tirthak Patel}
\IEEEauthorblockA{\textit{Rice University}\\Houston, TX, USA}
}

\maketitle

\begin{abstract}

We study the problem of \textit{probability distribution matching and sampling} on near-term quantum computers, aiming to construct parameterized circuits that generate samples from a target distribution while minimizing resource overhead. This task arises naturally in hybrid quantum–classical workflows, where measurement-driven objectives replace full state reconstruction, and is central to applications in generative modeling and variational inference. However, it remains challenging due to hardware noise, limited circuit depth, and a high-dimensional, non-convex parameter space.

We propose \sol{}, a surrogate-guided optimization framework based on Bayesian Optimization with tree-based models for scalable, domain-aware distribution matching. Our approach introduces a structured, layerwise decomposition aligned with the variational circuit architecture, enabling distributed and sample-efficient optimization within hybrid loops with theoretical convergence guarantees. Across representative distribution-matching tasks, \sol{} achieves up to \textit{2-3× lower total variation distance} while using \textit{40-60\% fewer gates} than prior approaches. These results demonstrate its effectiveness as a practical building block for end-to-end hybrid quantum sampling.

\end{abstract}

\section{Introduction}

Probability distribution matching and sampling is a core problem in quantum algorithm design, where the goal is to construct a low-depth, parameterized quantum circuit that reproduces the output distribution (measurement statistics) of a target process while minimizing resource overhead~\cite{joshi2025measuring,benedetti2025complement,hangleiter2023computational,nam2018automated,han2025enqode}. This task is especially critical in the context of near-term quantum hardware, which is constrained by short coherence times, limited gate fidelity, and strict circuit depth limits~\cite{preskill2018quantum,de2022survey}. Existing approaches often rely on domain-specific heuristics or gradient-based techniques~\cite{smith2021leap,younis2021qfast,khatri2019quantum} that either do not scale to high-dimensional parameter spaces or assume access to analytic gradients, which may not exist for circuits evaluated through noisy quantum measurements~\cite{murali2020software}. We therefore study distribution matching as a black-box optimization problem over a non-smooth, high-dimensional objective: the discrepancy between the output statistics of a parameterized circuit and a target distribution.

A natural approach is Bayesian Optimization (BO), which optimizes expensive black-box functions by constructing surrogate models~\cite{brochu2010tutorial,snoek2012practical,shahriari2015taking}. However, standard BO methods typically employ Gaussian Process (GP) surrogates, which scale poorly and require smoothness assumptions that do not hold in quantum optimization problems~\cite{wang2016bayesian}. In addition, GPs do not naturally capture the bounded distributions arising from quantum measurements and often oversmooth the non-smooth loss landscape. To address this, we propose \sol{}\footnote{This work is published at the IEEE International Conference on Quantum Computing and Engineering (QCE), 2026.}, a surrogate-guided framework for domain-aware distribution matching based on tree-based models following the spirit of~\cite{han2021high}, specifically gradient-boosted regression trees (GBRTs), which are better suited for the high-dimensional and non-smooth optimization landscape induced by quantum circuit outputs~\cite{patel2020experimental,patel2022quest}. \sol{} targets distribution matching based on output statistics, which is fundamental to tasks where measurement statistics, not full quantum amplitudes, determine success. These are precisely the near-term workloads limited by measurement access and noise~\cite{joshi2025measuring,benedetti2025complement,hangleiter2023computational}; our goal is to enable distributional alignment for sampling- and measurement-driven hybrid quantum–classical workflows, rather than general-purpose, tomographic unitary, or state synthesis.

We specifically show that these general-purpose methods are inefficient for our targeted domain. Beyond the surrogate choice, we introduce a structured decomposition of the parameter space that leverages the layered architecture of variational circuits~\cite{holmes_connecting_2022}. This layerwise decomposition yields a principled form of block-coordinate optimization: parameters within each layer are optimized in localized subspaces, while synchronization across layers ensures global convergence. This structure enables distributed, sample-efficient optimization and improves stability relative to random partitioning. We formalize the surrogate-guided distribution matching problem and provide theoretical guarantees under mild assumptions on the noise's stochasticity and the model's fidelity.

\noindent\textbf{Summary of this work's contributions:}

\begin{figure*}
    \centering
    \includegraphics[width=0.99\textwidth]{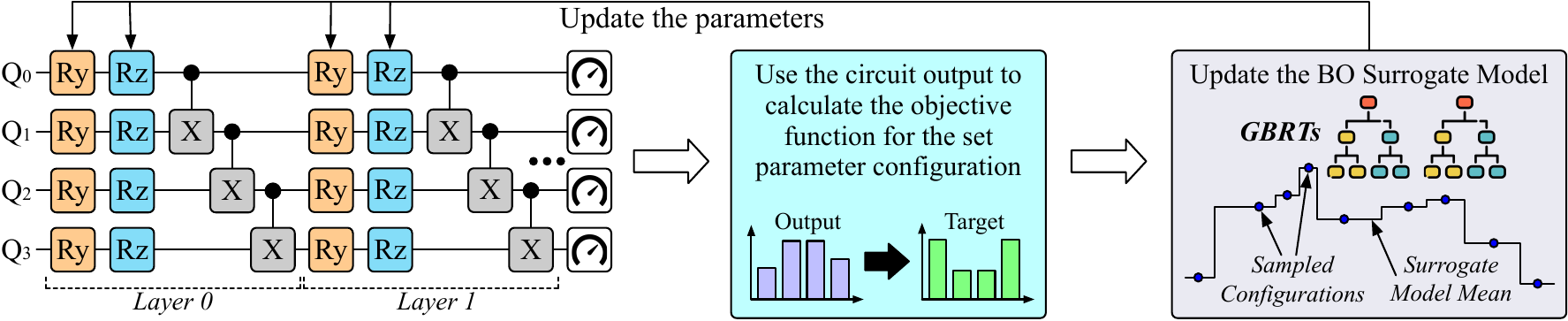}
    \caption{For \sol{}, we use BO with tree-based surrogates to update layered circuit parameters for probability distribution matching. We use a hardware-efficient ansatz to mitigate the effects of hardware noise.}
    \vspace{-5mm}
    \label{fig:bo}
\end{figure*}

\begin{itemize}[leftmargin=*]

\item We formulate domain-aware probability distribution matching as a black-box optimization problem with structured parameter spaces and identify the challenges of BO.

\item We propose a surrogate-guided framework, \sol{}, that leverages GBRT surrogates and a scalable, distributed subspace-optimization strategy based on circuit structure.

\item We provide convergence guarantees and analyze the impact of surrogate model fidelity and parameter decomposition on optimization performance for practical guidance.

\item We empirically validate the framework on widely-used quantum benchmarks, showing that our method achieves up to \textbf{2-3$\times$ lower total variation distance} while using \textbf{40-60\% fewer gates} than prior approaches.

\item \sol{}'s code and dataset are open-sourced at:\\ \textit{\url{https://github.com/positivetechnologylab/CircuitTree}}.

\end{itemize}
\section{Problem Setup}

Let $\Theta = [0,2\pi)^d$ denote the parameter space of an $n$-qubit parameterized quantum circuit $C(\bm{\theta})$ with $d$ parameterized gates, constructed from a fixed ansatz composed of $L$ layers of parameterized gates. Given an input state $\ket{\psi_0}$, the circuit induces an output probability distribution $p_{\bm{\theta}}$ over measurement outcomes in the computational basis.

Our goal is to match a target probability distribution $p^\star$ by optimizing the circuit parameters $\bm{\theta}$. The target distribution may arise from classical computation (e.g., analytic or data-driven distributions), prior quantum measurements, or domain-specific processes, depending on the application. This naturally leads to the following black-box optimization problem:

\begin{definition}[Distribution Matching Objective]
\[
\bm{\theta}^\star = \arg\min_{\bm{\theta} \in \Theta} \; \mathcal{L}(p_{\bm{\theta}}, p^\star),
\]
where $\mathcal{L}$ measures the discrepancy between the generated distribution $p_{\bm{\theta}}$ and the target distribution $p^\star$.
\end{definition}

We adopt the total variation distance (TVD)~\cite{oh2024classical,clark2024peephole,patel2021qraft} as the loss function, defined as the $\ell_1$ distance between probability vectors:
\[
\mathcal{L}(p_{\bm{\theta}}, p^\star) := \mathrm{TVD}(p_{\bm{\theta}}, p^\star) = \frac{1}{2} \sum_{x \in \{0,1\}^n} |p_{\bm{\theta}}(x) - p^\star(x)|.
\]

Each query to $\mathcal{L}$ is stochastic, as it is estimated from a finite number of measurements (shots). \rev{Our implementation utilizes sample-based estimates of TVD from limited measurement shots, which ensure consistent convergence and avoid exponential costs, which would otherwise be incurred with full state tomography-based distances.} Moreover, the objective is non-convex, non-differentiable, and highly non-smooth in general: small changes in $\bm{\theta}$ may propagate across layers and produce abrupt changes in output statistics~\cite{preskill2018quantum}. This motivates the development of surrogate models that can handle stochastic, discontinuous responses.

\begin{remark}
Unlike variational quantum algorithms, which optimize smooth cost functions derived from Hermitian observables, our objective arises directly from output distributions and is inherently non-smooth. This motivates the use of black-box optimization methods that do not rely on gradient information or smoothness~\cite{shahriari2015taking,luo2021hybrid,luo2024non}, and in particular surrogate models such as regression trees that naturally accommodate non-smoothness.
\end{remark}

To optimize this objective, we employ surrogate modeling. Let $f(\bm{\theta}) := \mathcal{L}(p_{\bm{\theta}}, p^\star)$ denote the true cost. The surrogate $\hat{f}_t$ is a learned approximation trained on observed evaluations:
\[
\mathcal{D}_t = \{ (\bm{\theta}_i, y_i) \}_{i=1}^t, \quad y_i = f(\bm{\theta}_i) + \xi_i
\]
where $\xi_i$ captures stochastic noise from measurement uncertainty or finite sampling. Evaluations on hardware may also include additional stochastic noise due to device imperfections such as thermal relaxation or depolarization~\cite{patel2020experimental,NEURIPS2019_f35fd567}.

\begin{definition}[Surrogate-Guided Optimization]
At each round $t$, the optimizer fits $\hat{f}_t$ on $\mathcal{D}_t$ and selects the next query point via an acquisition function $\alpha_t : \Theta \to \mathbb{R}$:
\[
\bm{\theta}_{t+1} = \arg\max_{\bm{\theta} \in \Theta} \alpha_t(\bm{\theta}; \hat{f}_t).
\]
\end{definition}

The acquisition function balances exploration and exploitation; common examples include Expected Improvement (EI) and Upper Confidence Bound (UCB)~\cite{brochu2010tutorial,shahriari2015taking}. The goal is to minimize $f(\bm{\theta})$ with as few queries as possible, yielding a shallow circuit that approximates the target distribution.

\begin{remark}
This problem departs from classical BO in key ways: (1) the function $f$ is distributional and highly non-smooth; (2) the parameter space $\Theta$ is structured by circuit layers and is only partially separable; and (3) the objective is defined relative to a fixed input state $\ket{\psi_0}$. These distinctions motivate both our surrogate choice (GBRT) and our structured layerwise optimization strategy.
\end{remark}  
\section{\sol{}'s Optimization Algorithm}
\label{sec:surrogates}

The core idea of our approach is to learn a surrogate model that approximates the true probability-matching loss $f(\bm{\theta}) := \text{TVD}(p_{\bm{\theta}}, p^\star)$ and to use this surrogate to guide parameter updates (Fig.~\ref{fig:bo}). Standard BO techniques often use Gaussian Process (GP) surrogates; however, GP-based models scale cubically with the number of observations, making them impractical in high-sample regimes~\cite{nicoli2024physics,benitez2024bayesian}. While GPs can be competitive for small datasets, they are also ill-suited for the highly non-smooth objectives that arise in probability distribution matching, such as minimizing TVD~\cite{williams2006gaussian}.

Instead, we employ Gradient Boosted Regression Trees (GBRTs), an ensemble of tree-based learners that capture sharp discontinuities~\cite{luo2021hybrid,hrluo_2019a} and perform well with limited data and at large scales, such as in quantum computing~\cite{nielsen2010quantum}. Using an ensemble rather than a single tree also enables the quantification of uncertainty needed for acquisition functions. Compared to other tree-based methods, GBRTs also allow us to demonstrate theoretical convergence and empirical outperformance, as discussed further in this section and Sec.~\ref{sec:experiments}.

\begin{definition}[Surrogate Model]
A surrogate model $\hat{f}_t : \Theta \to \mathbb{R}$ is a regression function trained to approximate $f$ using dataset $\mathcal{D}_t$. In our framework, $\hat{f}_t$ is a GBRT model composed of $M$ decision trees, each trained sequentially on residuals of the previous stage.

At each boosting step, a regression tree $h_{t}(\bm{\theta})$ is fit to the negative gradient of the loss $\mathcal{L}$ evaluated at the current prediction $\hat{f}_{t-1}$:
\[
h_{t}=\arg\min_{h}\sum_{i=1}^{t-1}\left[-\frac{\partial \mathcal{L}(y_{i},\hat{f}_{t-1}(\bm{\theta}_{i}))}{\partial\hat{f}_{t-1}(\bm{\theta}_{i})}\right]h(\bm{\theta}_{i}).
\]
The surrogate is updated by adding a scaled version of $h_{t}$:
\[
\hat{f}_{t}(\bm{\theta})=\hat{f}_{t-1}(\bm{\theta})+\nu \cdot h_{t}(\bm{\theta}),
\]
where $\nu$ is the learning rate controlling the contribution of each tree.
\end{definition}

\subsection{Acquisition Function and Optimization Strategy}

At each iteration $t$, the next query $\bm{\theta}_{t+1}$ is chosen by maximizing an acquisition function $\alpha_t : \Theta \to \mathbb{R}$ derived from the surrogate. We use the \emph{expected improvement} (EI) criterion:
\[
\alpha_t(\bm{\theta}) = \mathbb{E} \big[ \max(f^\star_t - \hat{f}_t(\bm{\theta}), 0 ) \big],
\]
where $f^\star_t = \min_{i \le t} y_i$ is the best observed value. In GBRTs, this expectation is approximated by quantile regression over ensemble predictions.

\begin{remark}
Unlike GPs, GBRTs do not natively provide posterior distributions. In \sol{}, we estimate uncertainty by combining (i) quantile-based predictions and (ii) diversity across tree paths in the ensemble, following the approach of~\cite{han2021high,meinshausen2006quantile}. This empirical posterior enables our use of the EI or UCB-style acquisition functions.
\end{remark}

\subsection{Layerwise Parameter Decomposition}

The parameter vector $\bm{\theta}$ is structured by circuit layers: each layer $\ell=1,\dots,L$ contains a subset $\bm{\theta}^{(\ell)}$. To exploit this structure, we introduce a distributed optimization strategy that partitions $\Theta$ into disjoint subspaces optimized independently, while others are fixed. To reduce the dimensionality of the search space, we explored splitting the parameter vector into subspaces and alternating optimization across them. At each iteration, one subspace is optimized while the others are held fixed, ensuring all parameters are eventually tuned. We refer to this process as \emph{subspace splitting}. A naive approach is to form subspaces by randomly grouping parameters. For the layered ansatz used in \sol{}, however, splitting by layers is more natural: each subspace corresponds to the parameters of one circuit layer. This improves interpretability, since subspaces align directly with the circuit's gate execution order.

\begin{figure*}[t]
\centering
\includegraphics[width=0.99\textwidth]{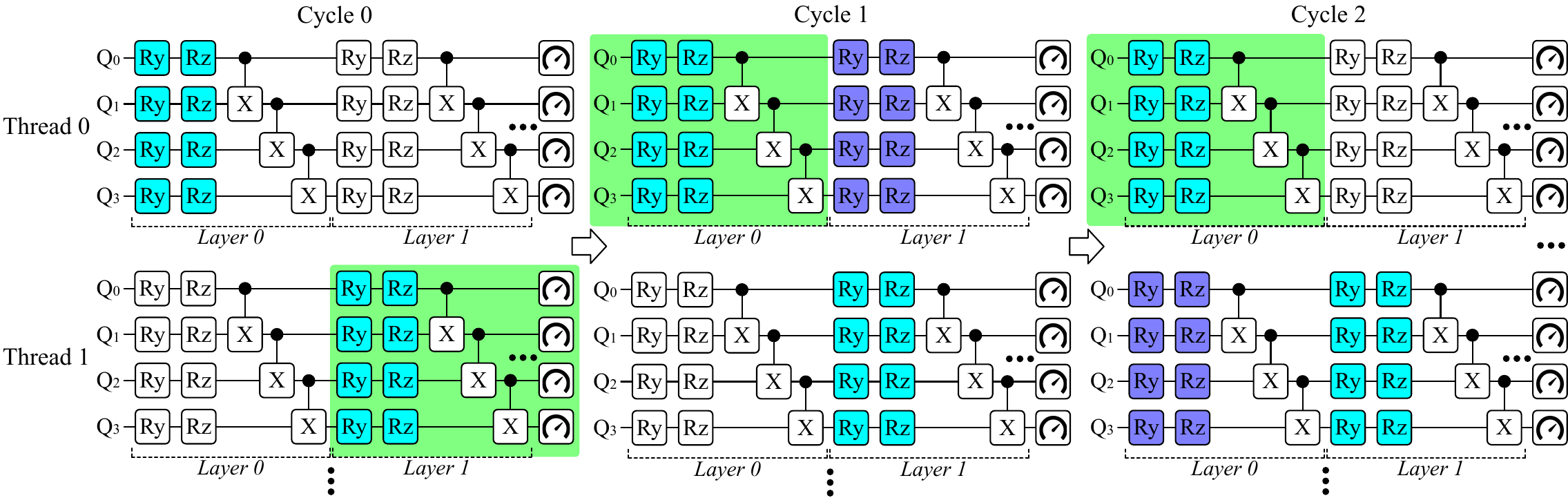}
\caption{Distributed subspace splitting: each thread optimizes one layer (blue). Improvements over the prior cycle (green) are propagated to the global parameter vector (purple) used by all threads.}
\label{fig:distributed}
\vspace{-5mm}
\end{figure*}

\subsection{Distributed Subspace Splitting}

While alternating subspaces improves trainability, we observed that the per-iteration progress within subspaces exceeded that of full-space optimization. To exploit this effect further, we developed a \emph{distributed subspace} method: each subspace is assigned to a separate thread, which trains its own surrogate and optimizes concurrently. After an initial warm-up over the full parameter space, each thread runs for a fixed number of iterations in its assigned subspace. Whenever a thread improves, it updates a shared global parameter vector, which is then synchronized across all threads. This scheme is illustrated in Fig.~\ref{fig:distributed}. This distributed approach combines the benefits of subspace optimization with global consistency.

\begin{definition}[Layerwise Decomposition]
Let $\Theta = \Theta^{(1)} \times \Theta^{(2)} \times \cdots \times \Theta^{(L)}$. For each layer $\ell$, a local surrogate $\hat{f}_t^{(\ell)}$ is trained on $\mathcal{D}_t$ restricted to $\Theta^{(\ell)}$.
\end{definition}

This yields a principled block coordinate optimization: (1) each surrogate operates in reduced dimensionality, improving sample efficiency; (2) layers can be optimized in parallel; and (3) barren plateaus are mitigated by restricting updates to local subspaces~\cite{holmes_connecting_2022}.

\begin{algorithm}[H]
\caption{\textsc{SurrogateMatch}$(p^\star, \mathcal{A})$}
\label{alg:synthesis}
\begin{algorithmic}[1]
\STATE Initialize $\bm{\theta}_0 \sim \text{Unif}(\Theta)$
\STATE Evaluate $y_0 = \text{TVD}(p({\bm{\theta}_0}), p^\star)$
\STATE Initialize dataset $\mathcal{D}_0 = {(\bm{\theta}_0, y_0)}$
\FOR{$t = 1$ to $T$}
\STATE Train GBRT surrogate $\hat{f}_t$ on $\mathcal{D}_{t-1}$
\FOR{each layer $\ell = 1,\dots,L$ \textbf{in parallel}}
\STATE Fix all $\bm{\theta}^{(j)}$ for $j \ne \ell$
\STATE Optimize $\alpha_t^{(\ell)}$ to get $\bm{\theta}^{(\ell)}_{t}$
\STATE Evaluate $y^{(\ell)}_t = \text{TVD}(p({\bm{\theta}_t}), p^\star)$
\STATE Update $\mathcal{D}_t \gets \mathcal{D}_{t-1} \cup {(\bm{\theta}_t, y^{(\ell)}_t)}$
\ENDFOR
\STATE Synchronize $\bm{\theta}_{t}$ across layers
\ENDFOR
\STATE \textbf{return} $\bm{\theta}_\text{best} = \arg\min_{(\bm{\theta}, y) \in \mathcal{D}_T} y$
\end{algorithmic}
\end{algorithm}

\subsection{Distributed Surrogate-Guided Optimization}

Our full algorithm is presented in Algorithm~\ref{alg:synthesis}. Each layer is optimized in parallel with periodic synchronization to ensure a globally consistent parameter set.

\begin{remark}
Layerwise decomposition with distributed surrogates improves sample efficiency, provides stability relative to random partitioning, and preserves convergence guarantees under mild assumptions. In addition, distribution matching only requires access to target output statistics in some applications (e.g., VQE); in others, such as classical data-driven sampling or generative tasks, the target distribution is directly available. When target statistics are required, their cost can be amortized across repeated sampling, making the approach practical in near-term hybrid quantum workflows.
\end{remark}
  
\section{Theoretical Analysis}
\label{sec:theory}

We now provide theoretical guarantees of convergence for \sol{}, our surrogate-guided approximate probability distribution-matching procedure. \rev{Our theoretical analysis builds upon standard proofs of Bayesian optimization convergence, with the novelty lying in adapting and proving these guarantees for non-Gaussian, tree-based surrogates under structured quantum parameter spaces, which, to our knowledge, is the first result of its kind.} Below we present a condensed analysis; full details are given in Appendix~\ref{sec:a1}. We begin with assumptions on the cost function and surrogate model class.

\begin{assumption}[Lipschitz Continuity]
\label{assumption:lipschitz}
The true loss $f: \Theta \to \mathbb{R}$ is $L$-Lipschitz w.r.t. the $\ell_2$ norm:
\[
|f(\bm{\theta}) - f(\bm{\theta}')| \le L \|\bm{\theta} - \bm{\theta}'\|_2 \quad \forall \bm{\theta}, \bm{\theta}' \in \Theta.
\]
\end{assumption}

\begin{assumption}[Bounded, Centered Noise]
\label{assumption:noise}
At step $t\ge1$ the algorithm queries $\bm{\theta}_{t}$ and observes
\[
y_{t}=f(\bm{\theta}_{t})+\xi_{t},\qquad\mathbb{E}[\xi_{t}]=0,\quad|\xi_{t}|\le\sigma\ \text{a.s.}
\]
\end{assumption}

\begin{assumption}[Variance Floor at Unexplored Points]
\label{assumption:variance}
For any unobserved $\tilde{\bm{\theta}}\notin\{\bm{\theta}_{i}\}_{i=1}^{t}$, at least one tree assigns $\tilde{\bm{\theta}}$ to an empty leaf determined by covariate splits. The ensemble predictive variance at $\tilde{\bm{\theta}}$ is bounded away from zero.
\end{assumption}

\begin{remark}
Assumption~\ref{assumption:lipschitz} is standard in BO analyses~\cite{shahriari2015taking}; although $f$ is globally non-smooth, local Lipschitzness suffices for regret bounds. Assumption~\ref{assumption:noise} is a simplification: while hardware noise is not strictly bounded, it is well-approximated by sub-Gaussian distributions with bounded variance due to error mitigation~\cite{preskill2021quantum,patel2020experimental,silver2023mosaiq}. Assumption~\ref{assumption:variance} follows prior tree-based BO work~\cite{luo2021hybrid,han2021high}, ensuring unexplored regions remain attractive under UCB/EI.
\end{remark}

To establish convergence, we first need to guarantee that unexplored regions do not collapse to zero variance under the surrogate.

\begin{lemma}[Predictive Variance at Unexplored Points]
\label{lemma:improvement}
Suppose $\tilde{\bm{\theta}}$ has never been queried up to round $t$. Then the ensemble variance satisfies
\[
s_t^2(\tilde{\bm{\theta}}) \geq \eta,
\]
where $\eta > 0$ depends only on past evaluations and the shrinkage parameter $\nu$.
\end{lemma}

Since unexplored regions remain attractive, the optimizer continues to spread queries throughout $\Theta$. We formalize this with the covering radius.

\begin{definition}[Covering Radius]
The covering radius after $t$ rounds is $\rho_t := \sup_{\bm{\theta} \in \Theta} \min_{1 \leq i \leq t} \|\bm{\theta} - \bm{\theta}_i\|_2$, the maximum distance from any $\bm{\theta}\in\Theta$ to its nearest sampled point.
\end{definition}

As the covering radius shrinks, every region of $\Theta$ is eventually explored. Combining this with Lipschitz continuity gives a guarantee for the \emph{incumbent} (best observed point), which is the standard output of Bayesian optimization.

\begin{definition}[Incumbent]
\label{def:incumbent}
Define the incumbent (best observed query) at round $t$ as
\[
\hat{\bm{\theta}}_t \in \arg\min_{1 \le i \le t} y_i.
\]
\end{definition}

\begin{theorem}[Convergence under Layered Distributed Optimization]
\label{thm:convergence}
Under Assumptions~\ref{assumption:lipschitz}--\ref{assumption:variance}, the incumbent $\hat{\bm{\theta}}_t$ produced by \textsc{SurrogateMatch} (Algorithm~\ref{alg:synthesis}) satisfies, for all $t\ge 1$,
\[
f(\hat{\bm{\theta}}_t) \le f^\star + L\rho_t + 2\sigma
\quad\text{a.s.},
\]
where $f^\star = \inf_{\bm{\theta} \in \Theta} f(\bm{\theta})$. Consequently,
\[
\limsup_{t \to \infty} \mathbb{E}\!\left[f(\hat{\bm{\theta}}_t)\right] \le f^\star + 2\sigma.
\]
If $\sigma = 0$ and $\rho_t \to 0$, then
\[
\lim_{t \to \infty} f(\hat{\bm{\theta}}_t) = f^\star
\quad\text{and hence}\quad
\lim_{t \to \infty} \mathbb{E}\!\left[f(\hat{\bm{\theta}}_t)\right] = f^\star.
\]
\end{theorem}

\begin{remark}[On rates]
The bound above is stated in terms of the covering radius $\rho_t$. If, in addition, the query sequence is constructed (or augmented) to be space-filling so that $\rho_t \le C\, t^{-1/d}$ for some constant $C$ (e.g., via an explicit low-discrepancy design or other covering guarantee), then in the noise-free case $\sigma=0$,
\[
\mathbb{E}\!\left[f(\hat{\bm{\theta}}_t)\right] - f^\star = \mathcal{O}(t^{-1/d}).
\]
Without such an additional space-filling condition, the theorem requires only $\rho_t \to 0$.
\end{remark}

\section{Discussion}
\label{sec:discussion}

\noindent\textbf{Surrogate Fidelity.} The accuracy of the surrogate model directly bounds the regret incurred at each iteration: lower surrogate error leads to tighter guarantees on expected improvement~\cite{shahriari2015taking}. Gaussian Processes assume smoothness and offer closed-form uncertainty estimates~\cite{snoek2012practical,williams2006gaussian}, which makes them effective in small-sample regimes but computationally prohibitive at scale due to cubic complexity in the number of evaluations~\cite{nicoli2024physics,benitez2024bayesian}. By contrast, tree-based surrogates such as GBRT~\cite{scikit-optimize,taieb2016forecasting} are agnostic to continuity and scale linearly with the number of samples, making them well suited for the non-smooth landscapes encountered in distribution matching and sampling from quantum circuits. Our analysis highlights the importance of ensemble-based acquisition heuristics to compensate for the lack of analytic posteriors~\cite{meinshausen2006quantile}.

\vspace{1mm}

\noindent\textbf{Structured Parameter Spaces.} Quantum circuits often follow layered, modular architectures~\cite{smith2021leap}, which induce a block structure in the parameter space. Our layerwise decomposition exploits this structure by reducing dimensionality at each step and enabling distributed surrogates, yielding a principled form of block coordinate optimization with convergence guarantees (Theorem~\ref{thm:convergence}). This aligns with prior results in distributed and multi-fidelity BO~\cite{swersky2013multi,kandasamy2015high}. Empirically, the structured updates also mitigate barren plateaus by focusing optimization on local subspaces~\cite{holmes_connecting_2022}. \rev{The mitigation arises not from altering gradient magnitudes but from restricting the optimization subspace through layerwise decomposition, which reduces parameter coupling/correlation and stabilizes optimization (explored empirically in Sec.~\ref{sec:experiments}).}

\vspace{1mm}

\begin{table}[t]
    \centering
    \caption{List of software libraries used for the implementation and evaluation of \sol{}.}
    \scalebox{0.86}{
    \begin{tabular}{cc|cc}
        \toprule
        \textbf{Software} & \textbf{Version} & \textbf{Software} & \textbf{Version} \\
        \midrule
        \texttt{python} & 3.12.3 & \texttt{scikit-optimize} & 0.10.2 \\
        \texttt{bqskit} & 1.1.2 & \texttt{qiskit-aer} & 0.14.2 \\
        \texttt{qiskit} & 1.1.0 & \texttt{qiskit-ibm-runtime} & 0.25.0 \\
        \texttt{SALib} & 1.5.0 & \texttt{scikit-learn} & 1.5.0 \\
        \bottomrule
    \end{tabular}}
    \label{tab:software}
    \vspace{-5mm}
\end{table}

\noindent\textbf{Expressivity vs. Trainability.} Highly expressive ans\"atze may require large parameter sets to approximate a target distribution~\cite{khatri2019quantum,holmes_connecting_2022}, but this increases dimensionality and degrades trainability. Layered decomposition offers a compromise: restricting updates to low-dimensional subspaces while preserving global convergence. This mirrors results in variational quantum learning~\cite{cerezo2021variational}, where expressivity often trades off against trainability due to barren plateaus. Our results suggest that architectural priors, parameter tying, and regularization can further improve trainability without sacrificing fidelity, consistent with recent advances in ML-inspired compilation~\cite{silver2022quilt,wang2022quantumnas}.

\vspace{1mm}

\noindent\textbf{Application Workflows and Hybrid Utility.} The distribution matching formulation naturally aligns with end-to-end hybrid quantum–classical workflows, where the objective is defined at the level of measurement statistics rather than full quantum states. Many near-term applications, including variational cost estimation, generative modeling, and data-driven quantum pipelines, operate on sampled outputs, making distributional alignment the relevant objective~\cite{joshi2025measuring,benedetti2025complement,hangleiter2023computational}. Our approach enables direct integration into hybrid loops, where circuit parameters are iteratively refined based on measured discrepancies with respect to a target distribution.

\vspace{1mm}

\noindent\textbf{Robustness and Noise-Aware Optimization.} Because the objective is defined over sampled measurement outcomes, our framework inherently operates under stochastic noise arising from finite-shot estimation and hardware imperfections. The use of tree-based surrogates allows the optimizer to remain stable under such noise, without relying on smoothness assumptions or analytic gradients. This is particularly important for near-term devices, where noise and variability are unavoidable, and aligns with practical deployment scenarios in which optimization must proceed under realistic hardware constraints~\cite{han2025enqode,wilson2020just}.
\section{Experimental and Analysis Methodology}
\label{sec:methodology}

\begin{figure*}[t]
    \centering
    \includegraphics[width=0.9\textwidth]{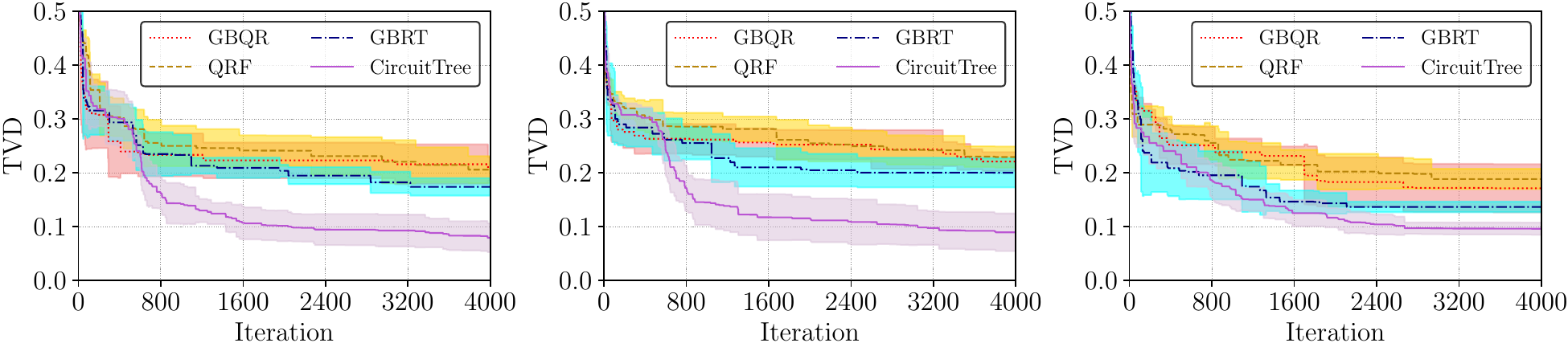}
    \caption{TVD during optimization of 3 different RQCs, using GBRT, QRF, and GBQR. GBRT significantly outperforms both QRF and GBQR in terms of TVD and runtime. \sol{}'s results with the final layered optimization design are shown.}
    \label{fig:surrogateconv}
    \vspace{-3mm}
\end{figure*}

\begin{figure*}[t]
    \centering
    \includegraphics[width=0.9\textwidth]{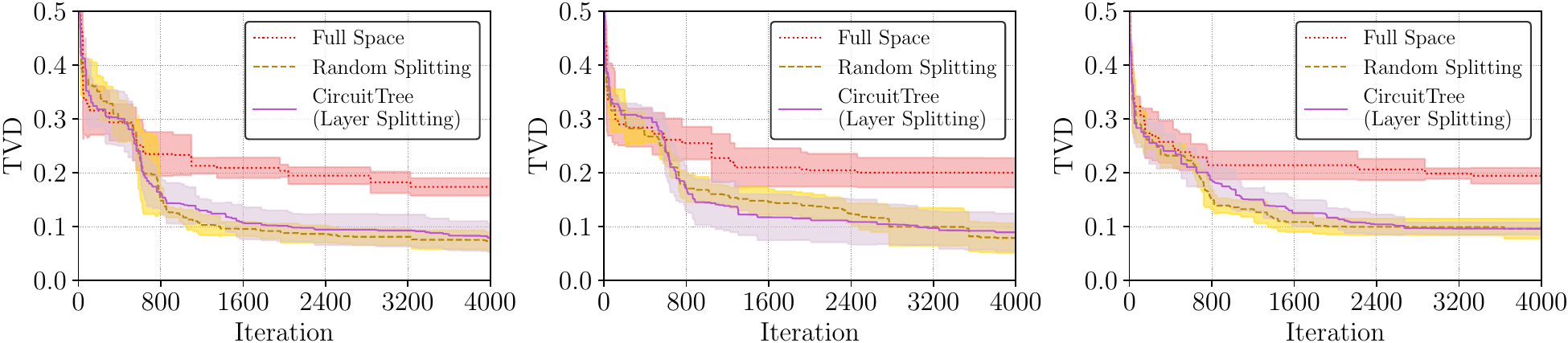}
    \caption{TVD during optimization of 3 different RQCs, using GBRT. Convergence is compared across full-space optimization, random subspace splitting, and layered splitting. \sol{} adopts layered splitting with distributed surrogate optimization.}
    \label{fig:layerwise_perf}
    \vspace{-5mm}
\end{figure*}

\subsection{Experimental Testbed Setup}

We run our synthesis experiments and classical processing tasks on our local computing cluster. The cluster consists of nodes with the AMD EPYC 7702P 64-core processor, x86\_64 architecture, and a 2.0 GHz clock. We spawn virtual machines (VMs) on these nodes with 8 cores, 32 GB of memory, and 32 GB of storage for each of our experiments, providing more than sufficient resources. The VMs are resource-bounded and not overprovisioned, ensuring that each experiment has exclusive access to the hardware resources assigned to it without any interference, which helps us provide accurate and consistent timing analysis. We run all of our hardware experiments on the \texttt{ibm\_nazca} quantum computer, a 127-qubit quantum computer with Eagle r3 architecture available via the IBM quantum cloud~\cite{castelvecchi2017ibm}. The computer has a median one-qubit gate error of $3.3\times{}10^{-4}$, a median two-qubit gate error of $1.2\times{}10^{-2}$, and a median measurement operation error of $2.3\times{}10^{-2}$. All circuits used in evaluation utilize four qubits.

\subsection{Software Framework Implementation}

Table~\ref{tab:software} provides a list of all the software used for the implementation and evaluation of \sol{}. All libraries and packages are Python-based. We use \texttt{scikit-optimize}~\cite{scikit-optimize} to perform BO, with models from the \texttt{scikit-learn} library~\cite{scikit-learn} as surrogates. We use the BQSKit library to run the state-of-the-art competitive synthesis framework~\cite{BQSKit2021} as a baseline for comparison. BQSKit~\cite{BQSKit2021} serves as an appropriate baseline as it represents a state-of-the-art quantum circuit synthesis framework that optimizes circuits under realistic hardware constraints. Although designed for circuit-level synthesis, it provides a strong point of comparison for our setting by producing circuits that implicitly induce target output distributions, allowing us to evaluate how effectively surrogate-guided distribution matching can achieve similar or improved sampling fidelity with reduced depth and gate complexity.

We use the \texttt{qiskit} library~\cite{aleksandrowiczqiskit} to create circuit instruction sets compatible with IBM quantum hardware, and \texttt{qiskit\_aer} to simulate quantum circuits and estimate output distributions during optimization. Our evaluation circuits are taken directly or modified from \texttt{QASMBench}~\cite{li2023qasmbench}, a benchmark suite of near-term quantum circuits.

We use \texttt{qiskit\_ibm\_runtime} to interface with the IBM quantum cloud and execute circuits on the \texttt{ibm\_nazca} quantum computer. When transpiling circuits, we use optimization level 0 to isolate the effect of our distribution matching approach from compiler-level optimizations. All experiments are run with 10,000 shots unless otherwise specified, and repeated five times to account for stochastic variability in both hardware and optimization. 

We implement repeated layers of the ansatz shown in Fig.~\ref{fig:bo} for \sol{}. The ansatz consists of parameterized Ry and Rz gates, followed by cascaded CX gates arranged in a linear topology. This structure enables the circuit to generate a rich family of output probability distributions while remaining hardware-compatible. The sparse connectivity reflects practical superconducting architectures, where dense connectivity introduces crosstalk and noise~\cite{dumitrescu2020benchmarking,wright2019benchmarking,ravi2021quantum}. 

By constraining the ansatz to this structure, \sol{} focuses on learning distributions that are realizable under hardware constraints, rather than exact circuit equivalence. This avoids the need for SWAP insertion and reduces the noise footprint. Each layer defines a compositional transformation of the output distribution, and we typically use 3–4 layers, which empirically provide a strong balance between expressivity and trainability for distribution matching. 

\subsection{Relevant Analysis Metrics}
\label{sec:metrics}

\noindent\textbf{Total Variation Distance (TVD).} The TVD is a widely-used metric to measure the difference between two probability distributions~\cite{oh2024classical,clark2024peephole,patel2022quest,patel2021qraft}. For a quantum system of $n$ qubits with $2^n$ output states, the TVD between two probability distributions $P_1$ and $P_2$ over these $2^n$ states can be measured as TVD $= \frac{1}{2}\sum_{i=0}^{2^n-1}\abs{p_1^i - p_2^i}$, where $p_1^i$ is the probability of observing state $i$ in distribution $P_1$ and $p_2^i$ is the probability of observing state $i$ in distribution $P_2$. This metric is used both during optimization and evaluation to quantify how well the learned circuit reproduces a target distribution. Unlike fidelity-based state metrics, TVD directly captures discrepancies in measurement outcomes, making it more appropriate for hybrid quantum workflows where only sampled outputs are observable.

\textbf{Synthesis Time.} This is the overall time required to optimize circuit parameters for a given target distribution. It reflects the classical overhead of the hybrid optimization loop. \textbf{Circuit Depth.} This measures the critical path length of the circuit and serves as a proxy for runtime and decoherence exposure~\cite{liu2020quantum,silver2023mosaiq,li2023qasmbench,wille2019mapping}. \textbf{Number of Gates.} We measure the total number of two-qubit CX gates, as these dominate error rates in near-term hardware~\cite{tannu2019not,ravi2021quantum,patel2020experimental,patel2020ureqa}. Together, these metrics capture the tradeoff between fidelity and efficiency.

\subsection{Algorithms Evaluated}
\label{sec:AlgoExpl}

We evaluate \sol{} across a set of representative workloads that define target probability distributions under different regimes relevant to near-term quantum applications.

\vspace{1mm}

\noindent\textbf{Randomly-Generated Quantum Circuits (RQC).} We use widely-used RQCs as a stress test for distribution matching, where the target distributions are induced by circuits with diverse and complex behaviors~\cite{Boixo2018}. In our setting, they serve as domain-agnostic targets to evaluate robustness and generality.

\vspace{1mm}

\noindent\textbf{Quantum State Preparation (QSP).} Amplitude embedding circuits define target probability distributions derived from classical data encoded into quantum amplitudes~\cite{Schuld2019,Grover2002,Mitarai2018}. We generate 10 such target distributions using \texttt{qiskit} state preparation routines~\cite{aleksandrowiczqiskit,shende2005synthesis}, and evaluate how well \sol{} can approximate them using a constrained ansatz. 

\vspace{1mm}

\noindent\textbf{Variational Quantum Eigensolver (VQE).} VQE defines target distributions implicitly through measurement statistics of parameterized quantum circuits that approximate ground states~\cite{peruzzo2014variational}. In our evaluation, these measurement distributions serve as targets for distribution matching, reflecting realistic hybrid workloads where objectives depend on sampled expectations. 


\section{Evaluation and Results}
\label{sec:experiments}


We aim to answer the following questions: \textbf{Q1} How do different surrogate models (GP vs. GBRT vs. Quantile Regression Forests (QRF)) compare in convergence speed, fidelity of distribution matching, and robustness? \textbf{Q2} What is the effect of layerwise distributed optimization on convergence time and stability? \textbf{Q3} How query-efficient is \sol{} in terms of quantum hardware measurements (shots), and how does it scale with ansatz depth and circuit width? \textbf{Q4} How does \sol{} compare with standard black-box optimizers under a fixed small budget of cost-function evaluations? \textbf{Q5} Can \sol{} reliably prepare domain-relevant samples on hardware, including those used in QSP and VQE?

\subsection{Surrogate Comparison (Q1)}

We first compare surrogate models for probability distribution matching. The candidates include: (1) Gaussian Processes (GPBO)~\cite{duong2022quantum,benitez2024bayesian,nicoli2024physics}, which provide probabilistic predictions and analytic uncertainty estimates but scale cubically in sample size and assume smoothness; (2) Gradient Boosted Quantile Regression (GBQR)~\cite{taieb2016forecasting}; and (3) Quantile Regression Forests (QRF)~\cite{meinshausen2006quantile}, both of which augment tree ensembles with explicit quantile modeling. All surrogates are embedded in the same Bayesian Optimization loop with Expected Improvement as the acquisition strategy. \textbf{Results.} Across three 3-layer RQCs, GBRT achieved the fastest convergence and lowest TVD (Fig.~\ref{fig:surrogateconv}). GPBO failed to finish within five days due to cubic scaling and the inability to capture sharp discontinuities. QRF and GBQR provided quantile-based uncertainty but introduced runtime overhead without improving fidelity. GBRT reached $\text{TVD} \leq 0.2$ with fewer evaluations and more than $2\times$ faster convergence, demonstrating robustness to non-smooth loss landscapes and practical suitability for near-term workloads.

\subsection{Layerwise Distributed Optimization (Q2)}

We next evaluate structured optimization strategies: (1) global surrogates trained over the full parameter space, (2) random subspace updates, and (3) our \emph{layerwise distributed optimization}, which assigns each circuit layer an independent surrogate updated in parallel. \textbf{Results.} Fig.~\ref{fig:layerwise_perf} shows that random subspaces improve over global surrogates but may introduce inconsistencies across layers, which may adversely affect convergence. Layerwise optimization achieved a $2.4\times$ reduction in convergence time and 50\% lower final TVD. The advantage grows as circuits deepen, where synchronization overhead is outweighed by locality-aware updates. Independent per-layer surrogates allow meaningful improvements without incurring global coordination costs at every step. These findings empirically validate our theoretical results (Theorem~\ref{thm:convergence}) and highlight the importance of exploiting ansatz structure for efficient probability distribution matching.

\begin{figure}[t]
    \centering
    \subfloat[Impact of Num. Layers on TVD]{\includegraphics[scale=0.44]{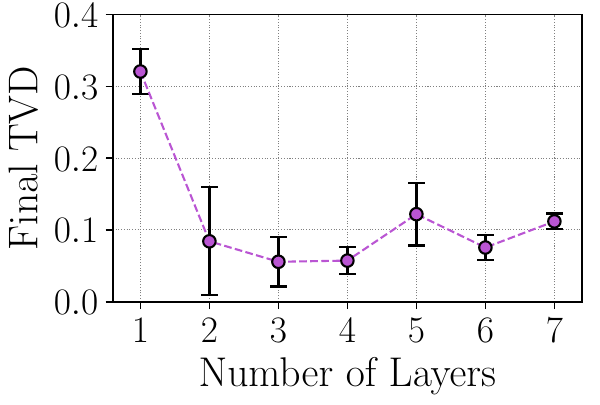}}
    \subfloat[Impact of Num. Shots on TVD]{\includegraphics[scale=0.44]{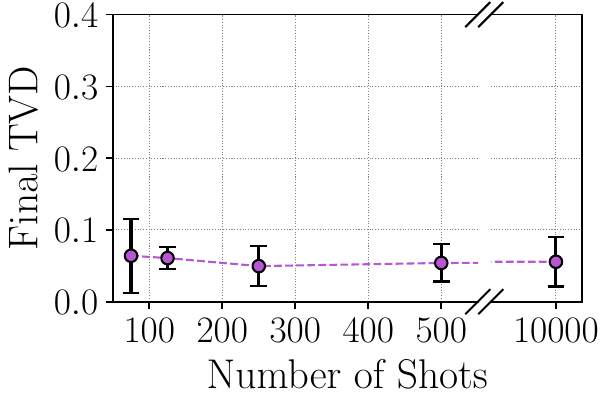}}
    \caption{Analyzing the impact of (a) the number of ans\"atze layers and (b) the number of measurement shots on the performance of \sol{} using VQE tasks.}
    \label{fig:scaling}
    \vspace{-2mm}
\end{figure}

\begin{table}[t]
\centering
\small
\caption{Reduced-budget simulator baselines under a fixed budget of 60 TVD objective evaluations. Each entry reports the mean $\pm$ standard deviation over 10 seeds.}
\label{tab:reduced_budget_baselines}
\begin{tabular}{l|cc|cc}
\hline
Task &
\multicolumn{2}{c|}{\sol{}} &
\multicolumn{2}{c}{GPBO} \\
 & TVD & Time (s) & TVD & Time (s) \\
\hline
VQE
& 0.40 $\pm$0.05 & 24.8 $\pm$1.7
& 0.41 $\pm$0.09 & 114 $\pm$32.9 \\
RQC0
& 0.33 $\pm$0.07 & 24.1 $\pm$1.2
& 0.35 $\pm$0.07 & 109 $\pm$38.8 \\
RQC1
& 0.31 $\pm$0.05 & 27.1 $\pm$2.3
& 0.33 $\pm$0.05 & 76.6 $\pm$9.5 \\
RQC2
& 0.29 $\pm$0.03 & 24.2 $\pm$1.3
& 0.32 $\pm$0.03 & 76.0 $\pm$16.6 \\
QSP0
& 0.17 $\pm$0.03 & 17.7 $\pm$0.2
& 0.20 $\pm$0.02 & 71.9 $\pm$10.5 \\
QSP1
& 0.23 $\pm$0.03 & 16.8 $\pm$0.6
& 0.24 $\pm$0.02 & 69.6 $\pm$12.0 \\
\hline
\end{tabular}
\vspace{-5mm}
\end{table}

\begin{figure*}[t]
    \centering
    \subfloat[\sol{} vs. BQSKit: TVD over Ideal Output]{\includegraphics[width=0.48\textwidth]{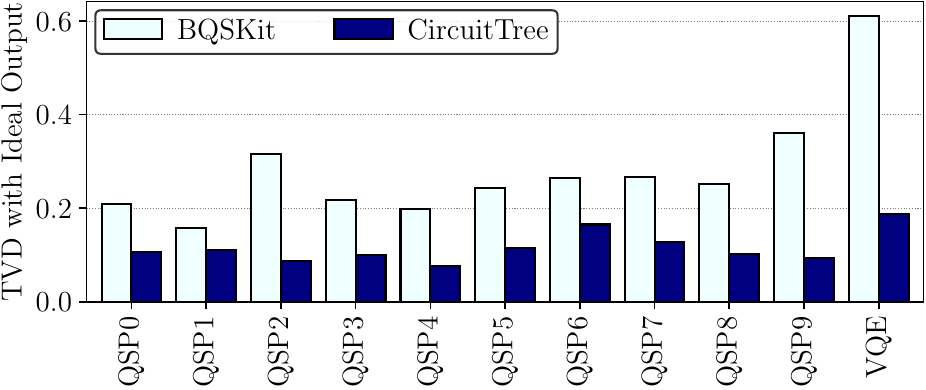}}
    \hfill
    \subfloat[\sol{} vs. BQSKit: Number of Gates]{\includegraphics[width=0.48\textwidth]{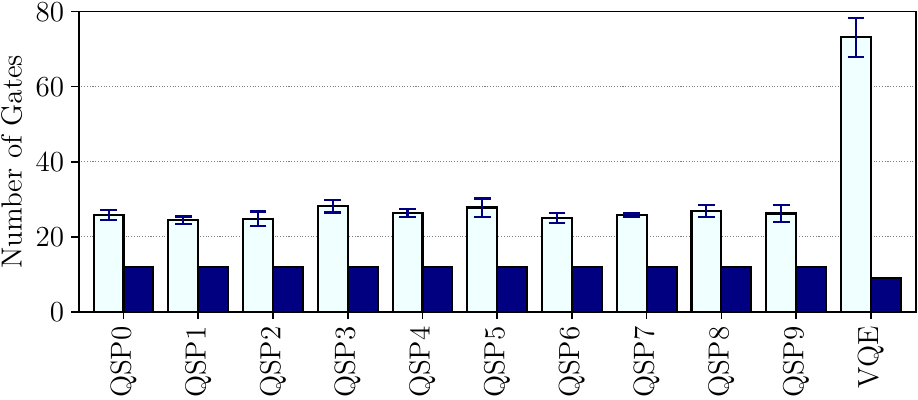}}
    \vspace{-2mm}
    \subfloat[\sol{} vs. BQSKit: Circuit Depth]{\includegraphics[width=0.48\textwidth]{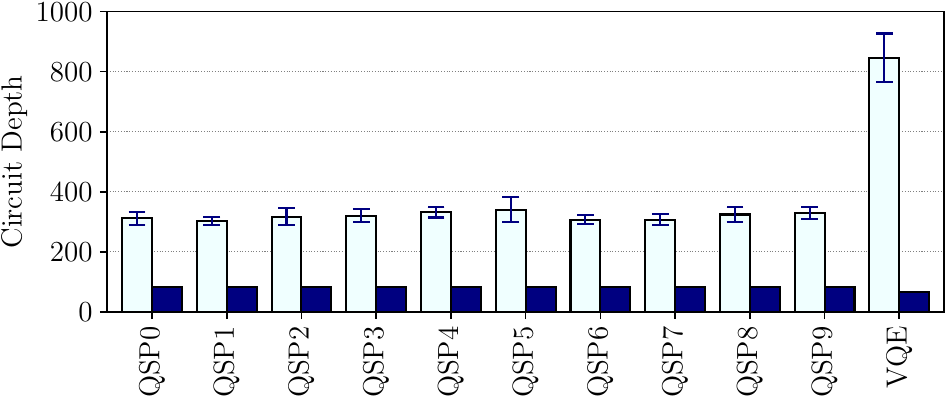}}
    \hfill
    \subfloat[\sol{} vs. BQSKit: Synthesis Time]{\includegraphics[width=0.48\textwidth]{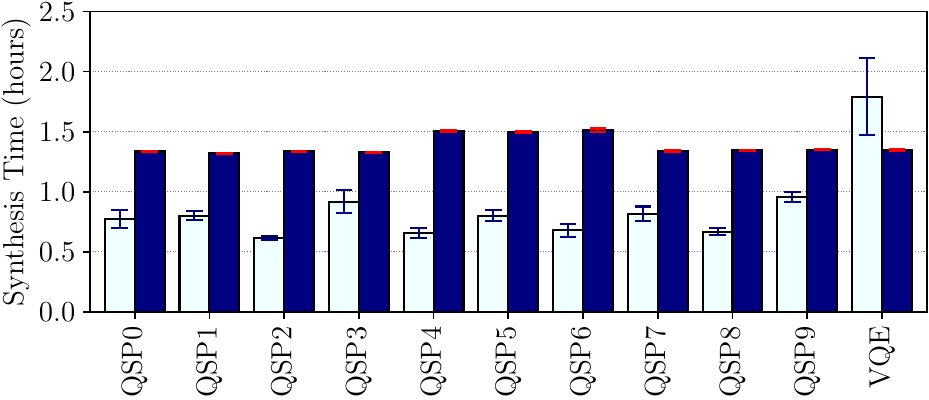}}
    \caption{Comparison of \sol{} and BQSKit on QSP and VQE workloads executed on IBM \texttt{ibm\_nazca}. \sol{} achieves higher fidelity with fewer gates and shallower depth, at the cost of increased but consistent classical runtime.}
    \label{fig:bqskit}
    \vspace{-5mm}
\end{figure*}

\subsection{Hardware-Efficient Scaling (Q3)}

We varied the number of ansatz layers (2--5) and the number of measurement shots (75 to 10,000) in the VQE sampling task to evaluate how these factors affect \sol{}'s fidelity. Each configuration was repeated across multiple seeds. \textbf{Results.} Fig.~\ref{fig:scaling} shows that three to four layers yield the best balance between expressivity and trainability: \sol{} consistently reached low TVD with minimal variance in this range. Two layers underfit the target distribution and exhibited unstable convergence, whereas five layers introduced overparameterization, degrading performance. Further, 250 shots were sufficient to achieve stable convergence. Using only 75 shots produced high variance and unreliable results, whereas increasing to 500 or even 10,000 shots offered no significant fidelity gains: a trend also observed for other circuits. These findings demonstrate that \sol{} achieves hardware-efficient matching with modest circuit depth and measurement overhead, making it well suited to near-term devices.

\subsection{Reduced-Budget Optimizer Baselines (Q4)}

Given the GPBO scaling problem discussed earlier, we now ask how it would perform under a reduced budget. \textbf{Results.} Table~\ref{tab:reduced_budget_baselines} summarizes the results under a fixed budget of 60 objective evaluations. \sol{} consistently outperforms GPBO across all six task–target instances in both final TVD and runtime. On VQE, \sol{} achieves $0.403 \pm 0.048$ compared to $0.410 \pm 0.091$ for GPBO while reducing runtime from $114$ s to $24.8$ s. Across the three RQCs, \sol{} improves TVD while reducing runtime from roughly $76$–$109$ s to $24$–$27$ s. On the QSP targets, \sol{} again outperforms GPBO in both fidelity and runtime. Overall, \sol{} outperforms GPBO even in the reduced-budget setting.

\subsection{Application Evaluation: QSP and VQE (Q5)}

We next evaluate \sol{} on application-relevant workloads executed on IBM’s \texttt{ibm\_nazca} quantum computer. This hardware study is complementary to the reduced-budget simulator baselines above and compares \sol{} against BQSKit across three metrics: (1) fidelity measured as TVD between hardware and ideal simulation, (2) circuit complexity (depth and two-qubit gate count), and (3) synthesis runtime.

\textbf{Results.} On real hardware, \sol{} reduced output error by up to 59\% compared to BQSKit (e.g., $0.12$ vs. $0.28$ for VQE), reduced two-qubit gate counts by 61\%, and shortened circuit depth by 78\%. These hardware-level improvements follow from surrogate-guided tuning of a fixed-depth ansatz, which ensures stable convergence and consistent circuit size. BQSKit, in contrast, produced variable-depth circuits with inconsistent fidelity. \sol{} also exhibited lower variance across runs on QSP tasks (0.02 vs. 0.06 standard deviation). While \sol{} incurred approximately $1.5\times$ higher classical runtime, this overhead was predictable, purely classical, and offset by fidelity gains and noise robustness. Importantly, trusted reference statistics were required only for VQE tasks and could be amortized across repeated uses of the distribution, making the approach practical in the near term.

\section{Related Work}
\label{sec:related}

\textbf{Bayesian Optimization and Surrogate Modeling.} Bayesian Optimization (BO) is a standard framework for optimizing expensive black-box functions~\cite{brochu2010tutorial,frazier2018tutorial}. Classical BO typically employs Gaussian Process surrogates due to their closed-form posterior updates and uncertainty quantification~\cite{snoek2012practical}. However, GP-based methods scale poorly in high dimensions and rely on smoothness assumptions that break down in the discontinuous loss landscapes induced by quantum measurements~\cite{wang2016bayesian}. Scalable alternatives have been proposed, including random forests~\cite{hutter2011sequential} and gradient-boosted trees~\cite{scikit-optimize}. Our contribution extends this line of work by analyzing tree-based surrogates in probability distribution matching and proving convergence guarantees under structured parameter spaces.

\textbf{Structured and Modular Optimization.} Decomposition strategies have been widely studied to improve sample efficiency in BO, including hierarchical models~\cite{swersky2013multi}, additive decompositions~\cite{kandasamy2015high}, and factorized acquisition rules~\cite{rolland2018high}. These often assume independence between subcomponents or rely on a known decomposition. In contrast, our approach exploits the explicit layered structure to define distributed surrogate subproblems. This is related to block-coordinate descent and regional-division methods~\cite{nesterov2012efficiency}, but differs in that the global objective is never fully evaluated; only distributional statistics from layered surrogates guide optimization.

\textbf{Quantum Circuit Synthesis and State Preparation.} Traditional circuit synthesis relies on algebraic, rule-based, or template-matching approaches~\cite{amy2013meet,nam2018automated,paradis2024synthetiq,Gidney2021,Kissinger2021,Miller2022,nicoli2024physics,tamiya2022stochastic,Amy2013}. More recent techniques include gradient-based variational optimization~\cite{khatri2019quantum} and probabilistic decomposition strategies~\cite{BQSKit2021,younis2021qfast}. These approaches typically assume access to gradients or explicit unitaries, both of which are impractical in near-term settings. Our work departs from this paradigm by explicitly targeting the domain of distribution matching, where gradients are unavailable and non-smoothness dominates, making optimization challenging.

\textbf{Machine Learning for Quantum Compilation.} There is increasing interest in applying ML to quantum compilation, transpilation, and general synthesis~\cite{czarnik2021error,zlokapa2023deep}. Most existing methods are empirical and heuristic, offering limited theoretical foundations. By contrast, our contribution provides the first provable convergence guarantees for probability distribution matching using non-Gaussian surrogates, leveraging ensemble tree models and structured optimization to achieve both scalability and theoretical rigor.

\section{Conclusion}
\label{sec:conclusion}

We presented \sol{}, a surrogate-guided framework for domain-aware probability distribution matching and sampling based on structured Bayesian Optimization. By combining tree-based surrogate models with a distributed, layerwise decomposition of the parameter space, our approach scales to high-dimensional, non-smooth objectives without relying on gradient information or full unitary access. We provided formal convergence guarantees under mild assumptions, and empirically validated the method’s efficacy on both simulated and real hardware. This work contributes to the growing field of structured hybrid quantum algorithm design, demonstrating that non-Gaussian surrogates can achieve both scalability and provable convergence in quantum settings.

\appendix

\section*{A. Proof for Theorem~\ref{thm:convergence}}
\label{sec:a1}

Let 
\[
\Theta \subset \mathbb{R}^d, \qquad 
D = \mathrm{diam}(\Theta) < \infty,
\]
\[
f : \Theta \to \mathbb{R}, \qquad 
f^\star = \inf_{\bm{\theta} \in \Theta} f(\bm{\theta}).
\]

We restate the main assumptions for clarity.

\paragraph{Assumption 4.1 (Lipschitz Continuity).}
The loss function $f$ is $L$-Lipschitz with respect to the $\ell_2$ norm:
\[
|f(\bm{\theta}) - f(\bm{\theta}')| \le L \|\bm{\theta} - \bm{\theta}'\|_2 
\quad \forall \bm{\theta}, \bm{\theta}' \in \Theta.
\]

\paragraph{Assumption 4.2 (Bounded, Centered Noise).}
At step $t \geq 1$, the algorithm queries $\bm{\theta}_t$ and observes
\[
y_t = f(\bm{\theta}_t) + \xi_t, \qquad 
\mathbb{E}[\xi_t] = 0, \quad 
|\xi_t| \le \sigma \ \text{almost surely.}
\]

\rev{Here $\xi_t$ models the stochastic noise in our evaluation of $f(\bm{\theta}_t)$
(for example, due to a finite number of measurement shots). 
The condition $\mathbb{E}[\xi_t]=0$ states that this estimator of $f(\bm{\theta}_t)$
is unbiased, while the bound $|\xi_t| \le \sigma$ provides a uniform control on the
magnitude of the noise.}

A gradient-boosted regression tree (GBRT) surrogate is fitted at each BO round $t$
using the dataset $D_t = \{(\bm{\theta}_i, y_i)\}_{i=1}^t$.
It is represented as an ensemble of $M_t$ regression trees
$h_{t,m} : \Theta \to \mathbb{R}$, combined with shrinkage parameter $0 < \nu \le 1$:
\[
\textstyle \hat{f}_t(\bm{\theta})
= \hat{f}_0(\bm{\theta}) + \nu \sum_{m=1}^{M_t} h_{t,m}(\bm{\theta}),
\]
where $\hat{f}_0$ is an initial constant predictor (e.g., the mean of $\{y_i\}$). At boosting stage $m$, we compute pseudo-residuals
\[
g_{t,m-1,i}
:= - \frac{\partial}{\partial \hat y}\,
\mathcal{L}\bigl(y_i, \hat{f}_{t,m-1}(\bm{\theta}_i)\bigr),
\qquad i = 1,\ldots,t,
\]
for a pointwise loss $\mathcal{L}(y,\hat y)$
(in our implementation, $\mathcal{L}(y,\hat y) = (y - \hat y)^2$ so that
$g_{t,m-1,i}$ are standard residuals).
The new tree $h_{t,m}$ is obtained by least-squares regression onto these
pseudo-residuals:
\[
h_{t,m}
= \arg\min_{h} \sum_{i=1}^t \bigl(g_{t,m-1,i} - h(\bm{\theta}_i)\bigr)^2.
\]
We then update the ensemble as
\[
\hat{f}_{t,m}(\bm{\theta})
= \hat{f}_{t,m-1}(\bm{\theta}) + \nu\, h_{t,m}(\bm{\theta}),
\]
and set $\hat{f}_t := \hat{f}_{t,M_t}$.

The empirical mean and variance across trees at round $t$ are
\[
\textstyle \mu_t(\bm{\theta})
= \frac{1}{M_t}\sum_{m=1}^{M_t} h_{t,m}(\bm{\theta}),
\]
\[
\textstyle s_t^2(\bm{\theta})
= \frac{1}{M_t}\sum_{m=1}^{M_t}
\bigl(h_{t,m}(\bm{\theta}) - \mu_t(\bm{\theta})\bigr)^2.
\]

The algorithm selects query points via a confidence-bound rule for minimization:
\begin{equation}
\bm{\theta}_{t+1} = 
\arg\min_{\bm{\theta} \in \Theta} \Bigl\{ \mu_t(\bm{\theta}) - \kappa_t s_t(\bm{\theta}) \Bigr\},
\label{eq:UCB}
\end{equation}
where $\kappa_t > 0$ is an exploration multiplier.  
Empirically, we use expected improvement (EI) as the acquisition function; for the analysis we work with the confidence-bound rule~\eqref{eq:UCB}, which plays a similar exploration--exploitation role.

\begin{definition}[Covering Radius]
The covering radius at round $t$ is
\[
\rho_t := \sup_{\bm{\theta} \in \Theta} \min_{1 \leq i \leq t} \|\bm{\theta} - \bm{\theta}_i\|_2.
\]
\end{definition}

\subsection{Lower Bound on Ensemble Variance at Unexplored Points}

\begin{lemma}
Fix $t \geq 1$. Suppose $\tilde{\bm{\theta}} \in \Theta$ has never been queried, i.e.\ $\tilde{\bm{\theta}} \neq \bm{\theta}_i$ for all $1 \le i \le t$.  
Assume that at least one tree assigns $\tilde{\bm{\theta}}$ to an empty leaf, i.e.\ a region of the partition containing no training points.  
Then the ensemble variance satisfies
\[
s_t^2(\tilde{\bm{\theta}}) \geq \eta,
\]
where $\eta > 0$ depends only on past evaluations and the shrinkage
parameter $\nu$, and the ensemble size sequence $\{M_u\}_{u\le t}$.
\end{lemma}

\paragraph{Proof.}
\begin{enumerate}
\item For each tree $h_m$, let $\ell_m(\bm{\theta})$ denote the leaf containing $\bm{\theta}$.  
Define $A_m(\bm{\theta}) = \{ i \le t : \bm{\theta}_i \in \ell_m(\bm{\theta})\}$, i.e.\ the indices of training points in the same leaf.  
The leaf prediction is
\[
h_m(\bm{\theta}) :=
\begin{cases}
\displaystyle \frac{1}{|A_m(\bm{\theta})|} \sum_{i \in A_m(\bm{\theta})} r_{m,i}, & \text{if } |A_m(\bm{\theta})|>0,\\[0.6em]
0, & \text{if } |A_m(\bm{\theta})|=0.
\end{cases}
\]
where $r_{m,i}$ is the residual for sample $i$ at tree $m$.

\item Square-loss boosting fits each tree $h_m$ to residuals $r_{m,i} = y_i - \hat{f}_{m-1}(\bm{\theta}_i)$.  
Least-squares fitting ensures
\[
\textstyle \frac{1}{t}\sum_{i=1}^t r_{m^\star,i}^2 
= \min_c \frac{1}{t}\sum_{i=1}^t (r_{m^\star,i} - c)^2.
\]
Taking $c = \bar{r}_{m^\star} = \tfrac{1}{t}\sum_i r_{m^\star,i}$ yields
\[
\textstyle s_{\mathrm{res}}^2 := \frac{1}{t}\sum_{i=1}^t (r_{m^\star,i} - \bar{r}_{m^\star})^2 > 0,
\]
since residuals cannot all be identical under bounded but varying noise.

\item For the tree $m^\star$ with an empty leaf at $\tilde{\bm{\theta}}$, we have $h_{m^\star}(\tilde{\bm{\theta}}) = 0$.  
Using variance decomposition across trees,
\[
s_t^2(\tilde{\bm{\theta}}) \;\ge\; \frac{\nu^2}{M_t} s_{\mathrm{res}}^2.
\]
Since $M_t \le M_{\max}$, we conclude
\[
s_t^2(\tilde{\bm{\theta}}) \;\ge\; \frac{\nu^2}{M_{\max}} s_{\mathrm{res}}^2 =: \eta > 0.
\]
\end{enumerate}
\qed

\subsection{Density of Queries in \texorpdfstring{$\Theta$}{Theta}}

\begin{lemma}
Fix $r > 0$ and $\tilde{\bm{\theta}} \in \Theta$.  
There exists a finite index $t_r(\tilde{\bm{\theta}})$ such that
\[
\|\bm{\theta}_{t_r(\tilde{\bm{\theta}})} - \tilde{\bm{\theta}}\|_2 \le r.
\]
\end{lemma}

\paragraph{Proof.}
Let $B(\tilde{\bm{\theta}}, r)$ denote the open $r$-ball around $\tilde{\bm{\theta}}$.  
Suppose, for contradiction, that no $\bm{\theta}_i$ with $i \le t$ lies in $B(\tilde{\bm{\theta}}, r)$.  
Then Lemma~1 implies $s_{i-1}(\bm{\theta}) \ge \eta$ for all $\bm{\theta} \in B(\tilde{\bm{\theta}}, r)$.  

As more points are sampled, the predictive variance at already-sampled points 
$s_{i-1}(\bm{\theta}_{i-1})$ decreases; assuming in particular that 
$s_{i-1}(\bm{\theta}_{i-1}) \to 0$, choose $\kappa_{i-1}$ large enough 
(e.g., with $\kappa_t \to \infty$ slowly, such as $\kappa_t = \sqrt{\log t}$) that
\[
\mu_{i-1}(\bm{\theta}_{i-1}) - \kappa_{i-1} s_{i-1}(\bm{\theta}_{i-1}) 
>
\]
\[
\inf_{\bm{\theta} \in B(\tilde{\bm{\theta}}, r)} 
\Bigl\{ \mu_{i-1}(\bm{\theta}) - \kappa_{i-1} s_{i-1}(\bm{\theta}) \Bigr\}.
\]
By the UCB rule in \eqref{eq:UCB}, the next query point lies in $B(\tilde{\bm{\theta}}, r)$, contradicting the assumption.  
Hence such a $t_r(\tilde{\bm{\theta}})$ exists.  
Applying a Borel–Cantelli argument to a countable basis of rational balls implies $\lim_{t \to \infty} \rho_t = 0$, i.e.\ the query sequence is dense in $\Theta$. \qed

\subsection{Simple regret bound from the covering radius}

We bound the quality of the \emph{incumbent} (best observed query) using only Lipschitz continuity
and covering radius.

\begin{definition}[Incumbent and incumbent simple regret]
For $t\ge 1$, define an incumbent index
\[
\hat{\imath}_t \in \arg\min_{1 \le i \le t} y_i,
\]
and the corresponding incumbent point
\[
\hat{\bm{\theta}}_t := \bm{\theta}_{\hat{\imath}_t}.
\]
Define its (true) simple regret by
\[
r_t^{\mathrm{inc}} := f(\hat{\bm{\theta}}_t) - f^\star.
\]
\end{definition}

\begin{lemma}[Lipschitz bound for the incumbent]
\label{lem:incumbent_bound}
Assume Assumptions 4.1 and 4.2. For every $t \ge 1$,
\[
r_t^{\mathrm{inc}} \le L\rho_t + 2\sigma
\quad\text{almost surely.}
\]
In particular,
\[
\mathbb{E}[r_t^{\mathrm{inc}}] \le L\,\mathbb{E}[\rho_t] + 2\sigma.
\]
\end{lemma}

\paragraph{Proof.}
Fix any $\varepsilon>0$ and choose $\bm{\theta}^\star_\varepsilon \in \Theta$ such that
\[
f(\bm{\theta}^\star_\varepsilon) \le f^\star + \varepsilon,
\]
which exists by the definition of $f^\star=\inf_{\bm{\theta}\in\Theta} f(\bm{\theta})$.
By the definition of the covering radius $\rho_t$, there exists an index
$i^\star_\varepsilon(t)\in\{1,2,\dots,t\}$ such that
\[
\|\bm{\theta}_{i^\star_\varepsilon(t)}-\bm{\theta}^\star_\varepsilon\|_2 \le \rho_t.
\]
By Lipschitz continuity (Assumption 4.1),
\[
f(\bm{\theta}_{i^\star_\varepsilon(t)}) \le f(\bm{\theta}^\star_\varepsilon) + L\rho_t
\le f^\star + \varepsilon + L\rho_t.
\]
Since $\hat{\imath}_t$ minimizes the observed values $y_i$ over $i\le t$,
\[
y_{\hat{\imath}_t} \le y_{i^\star_\varepsilon(t)}.
\]
Using the bounded noise model $y_i = f(\bm{\theta}_i) + \xi_i$ with $|\xi_i|\le\sigma$
(Assumption 4.2), we have
\[
f(\hat{\bm{\theta}}_t) \le y_{\hat{\imath}_t} + \sigma
\le y_{i^\star_\varepsilon(t)} + \sigma
\le f(\bm{\theta}_{i^\star_\varepsilon(t)}) + 2\sigma
\le f^\star + \varepsilon + L\rho_t + 2\sigma.
\]
Rearranging yields $r_t^{\mathrm{inc}} \le L\rho_t + 2\sigma + \varepsilon$.
Since $\varepsilon>0$ is arbitrary, we conclude $r_t^{\mathrm{inc}} \le L\rho_t + 2\sigma$. \qed

\begin{remark}[About rates]
A classical covering argument implies that for any diameter-$D$ subset of $\mathbb{R}^d$,
there exists a set of $t$ points whose covering radius is at most
\[
\rho_t \le \left(\frac{C_d D^d}{t}\right)^{1/d},
\qquad
C_d = \frac{\pi^{d/2}}{\Gamma(1+d/2)}.
\]
This bound describes what is achievable by a \emph{space-filling} design (or the optimal
placement of $t$ points). If the query sequence is constructed (or augmented) to satisfy such a bound, then the
noise-free incumbent regret obeys $r_t^{\mathrm{inc}} = O(t^{-1/d})$ by
Lemma~\ref{lem:incumbent_bound}.
\end{remark}

\subsection{Theorem Statement}

From the density argument above, $\rho_t \to 0$ as $t\to\infty$.
Applying Lemma~\ref{lem:incumbent_bound} yields
\[
\limsup_{t \to \infty} \mathbb{E}\!\left[f(\hat{\bm{\theta}}_t)\right] \le f^\star + 2\sigma,
\]
and in the noise-free case $\sigma = 0$,
\[
f(\hat{\bm{\theta}}_t) \le f^\star + L\rho_t \to f^\star,
\]
hence $\lim_{t\to\infty} f(\hat{\bm{\theta}}_t)=f^\star$.
This proves Theorem~\ref{thm:convergence}.

\section*{Acknowledgement}

We thank the anonymous reviewers for their helpful comments, which improved this work. This work was supported by Rice University, Santa Clara University, the Rice University George R. Brown School of Engineering and Computing, the Rice University Department of Computer Science, the DOE Quantum Testbed Finder Award DE-SC0024301, the Ken Kennedy Institute, and the Rice Quantum Initiative within the Smalley-Curl Institute. Hengrui Luo was supported by the U.S. Department of Energy under Contract DE-AC02-05CH11231 and the U.S. National Science Foundation under Grant NSF-DMS 2412403. Generative AI tools were used for code and text editing, and all edits were reviewed and verified by the authors. We acknowledge the use of IBM Quantum services. The views expressed are those of the authors and do not necessarily reflect those of IBM or the IBM Quantum team.

\balance

\bibliographystyle{unsrt}
\bibliography{main}

\end{document}